\newcommand{\beq}{\begin{equation}}
\newcommand{\eeq}{\end{equation}}
\newcommand{\beqa}{\begin{eqnarray}}
\newcommand{\eeqa}{\end{eqnarray}}
\newcommand{\ba}{\begin{array}}
\newcommand{\ea}{\end{array}}

\documentclass[prc,twocolumn,showpacs,preprintnumbers,
amsmath,amssymb,floatfix]{revtex4}
\usepackage{hyperref}
\usepackage{epsfig}

\begin{document}

\title{Condensate fraction in neutron matter} 

\author{Luca Salasnich}
\affiliation{Dipartimento di Fisica ``Galileo Galilei'' and CNISM, 
Universit\`a di Padova, Via Marzolo 8, 35131 Padova, Italy}

\date{\today}

\begin{abstract}
We study the Bose-Einstein condensation of fermionic pairs in 
the uniform neutron matter by using the concept 
of the off-diagonal long-range order of the two-body 
density matrix of the system. 
We derive explicit formulas for the condensate density $\rho_{con}$
and the condensate fraction $\rho_{con}/\rho$ in terms 
of the scaled pairing energy gap $\Delta/\epsilon_F$ where 
$\epsilon_F$ is the Fermi energy. We calculate the condensate fraction 
$\rho_{con}/\rho$ as a function of the density $\rho$ by using 
previously obtained results for the pairing gap $\Delta$. We find 
the maximum condensate fraction $(\rho_{con}/\rho)_{max}= 0.42$ 
at the density $\rho=5.3\cdot 10^{-4}$ fm$^{-3}$, which corresponds 
to the Fermi wave number $k_F= 0.25$ fm$^{-1}$.  
\end{abstract}


\maketitle

The concept of off-diagonal long-range order (ODLRO) was introduced by 
Penrose and Onsager \cite{penrose} to determine the condensate 
fraction of a generic bosonic many-body system. Yang \cite{yang} 
extended this powerful idea to the case of fermionic 
many-body systems. 
It is now established that at zero temperature 
the condensate fraction of liquid $^4$He is below $10\%$ 
\cite{ceperly}, while 
for dilute and ultracold bosonic alkali-metal atoms it can 
reach $100\%$ \cite{review}. Few years ago the condensate fraction 
has been calculated for fermionic atoms \cite{sala-odlro,ortiz} in 
the crossover from the Bardeen-Cooper-Schrieffer (BCS) 
state of Cooper Fermi pairs to the Bose-Einstein condensate (BEC) 
of molecular dimers at zero temperature. 
These theoretical predictions are in quite good agreement 
with the data obtained in two experiments \cite{zwierlein,ueda} 
with ultra-cold two hyperfine component Fermi vapors of $^6$Li atoms. 
Very recently, the condensate fraction of dilute neutron matter, 
which is predicted to fill the crust of neutron stars \cite{star},
has been calculated by Wlazlowski and Magierski \cite{magierski1,magierski2}. 
Unfortunately, these finite-temperature Path Integral Monte Carlo 
simulations are performed only at $\rho=0.003$ fm$^{-3}$, 
and their trend suggests a condensate fraction 
of about $35\%$ at zero temperature \cite{magierski2}. 

In this paper we present a zero-temperature systematic study 
of the condensate fraction of neutron matter as a function of the 
density $\rho$. First we derive explicit analytical formulas 
for the condensate density $\rho_{con}$ 
and the condensate fraction $\rho_{con}/\rho$ 
as a function of the pairing energy gap $\Delta$ 
in units of the Fermi energy $\epsilon_F$. With the help of these 
expressions we calculate the condensate fraction 
$\rho_{con}/\rho$ vs $\rho$ by using 
previously obtained results for the paring gap $\Delta$ 
\cite{matsuo,carlson}. 

The shifted Hamiltonian of the uniform neutron matter is given by  
\beqa 
{\hat H}' &=& 
\int d^3{\bf r} \
\sum_{\sigma=\uparrow , \downarrow}
\ {\hat \psi}^+_{\sigma}({\bf r}) 
\left(-{\hbar^2\over 2 m}\nabla^2- \mu \right) {\hat \psi}_{\sigma}({\bf r}) 
\label{ham} 
\\
&+&
\int d^3{\bf r} \ d^3{\bf r}' \  
{\hat \psi}^+_{\uparrow}({\bf r})\ {\hat \psi}^+_{\downarrow}({\bf r}')\ 
V({\bf r}-{\bf r}')\ 
{\hat \psi}_{\downarrow}({\bf r}')\ {\hat \psi}_{\uparrow}({\bf r}) \; , 
\nonumber
\eeqa 
where ${\hat \psi}_{\sigma}({\bf r})$ is the field operator 
that annihilates a neutron of spin $\sigma$ 
in the position ${\bf r}$, while ${\hat \psi}_{\sigma}^+({\bf r})$ 
creates a neutron of spin $\sigma$ in ${\bf r}$. Here $V({\bf r}-{\bf r}')$ 
is the nucleon-nucleon potential in the $^1$S$_0$ channel. 
The ground-state average of the number of neutrons reads 
\beq 
N=\int d^3{\bf r} \ \sum_{\sigma=\uparrow,\downarrow} \ 
\langle \, 
{\hat \psi}^+_{\sigma}({\bf r})\ {\hat \psi}_{\sigma}({\bf r}) 
\, \rangle = \int d^3{\bf r}\ \langle \, {\hat \rho}({\bf r})\, \rangle \; . 
\label{def-n}
\eeq
This total number $N$ is fixed by the chemical potential $\mu$ 
which appears in Eq. (\ref{ham}). 

As shown by Yang \cite{yang}, for a Fermi system ODLRO means 
that the two-body density matrix factorizes as follows 
\beqa
\langle \, 
{\hat \psi}^+_{\uparrow}({\bf r}_1')\ {\hat \psi}^+_{\downarrow}({\bf r}_2') 
\ {\hat \psi}_{\downarrow}({\bf r}_1)\ {\hat \psi}_{\uparrow}({\bf r}_2) \, 
\rangle 
\nonumber 
\\
= \langle \,  
{\hat \psi}^+_{\uparrow}({\bf r}_1')\ 
{\hat \psi}^+_{\downarrow}({\bf r}_2') \, \rangle \ 
\langle \, 
{\hat \psi}_{\downarrow}({\bf r}_1)\ {\hat \psi}_{\uparrow}({\bf r}_2) \, 
\rangle \;  
\eeqa 
in the limit wherein both unprimed coordinates approach
an infinite distance from the primed coordinates. 
The largest eigenvalue $N_0$ of the two-body density matrix 
gives the number of neutron pairs which have their center of mass 
with zero linear momentum. This condensed number of pairs 
is given by \cite{sala-odlro,ortiz}
\beq
N_{con} = 2 \int d^3{\bf r}_1 \; d^3{\bf r}_2 \; | \langle \, 
{\hat \psi}_{\downarrow}({\bf r}_1) \  
{\hat \psi}_{\uparrow}({\bf r}_2) \, \rangle |^2 \; .
\label{def-n0}
\eeq 

Within the Hartree-Fock-Bogoliubov approach \cite{fetter}, 
the shifted Hamiltonian density (\ref{ham}) 
can be diagonalized by using the Bogoliubov-Valatin  
representation of the field operator 
${\hat \psi}_{\sigma}({\bf r})$ 
in terms of the anticommuting quasi-particle 
Bogoliubov operators ${\hat b}_{{\bf k}\sigma}$ 
with amplitudes $u_{\bf k}$ and $v_{\bf k}$ and 
quasi-particle energy $E_{\bf k}$. 
After minimization of the resulting quadratic Hamiltonian 
one finds familiar expressions for these quantities \cite{fetter}: 
\beq
E_{\bf k}=\left[(\tilde{\epsilon}_{\bf k}-\mu)^2 + 
\Delta_{\bf k}^2\right]^{1/2}
\eeq 
and
\beq 
u_{\bf k}^2 = {1\over 2} \left( 1 + \frac{\tilde{\epsilon}_{\bf k} 
- \mu }{E_{\bf k}} \right) \; , \quad
v_{\bf k}^2 = {1\over 2} \left( 1 - \frac{\tilde{\epsilon}_{\bf k} 
- \mu }{E_{\bf k}} \right) \, ,  
\label{uev}
\eeq 
where $\tilde{\epsilon}_{\bf k}$ is the single-particle 
energy and $\Delta_{\bf k}$ is 
the pairing gap, which satisfies the integral equation 
\beq
\Delta_{\bf q} = {1\over 2} \sum_{\bf k} 
\langle {\bf q},-{\bf q}|V|{\bf k},-{\bf k}\rangle 
\ {\Delta_{\bf k} \over E_{\bf k}} \; . 
\label{delta}
\eeq 
In the Hartree-Fock-Bogoliubov approach \cite{fetter} 
the single-particle energy $\tilde{\epsilon}_k$ is given by 
$\tilde{\epsilon}_{\bf k} = \epsilon_{\bf k}+U_{\bf k}$, 
where $\epsilon_{\bf k}=\hbar^2k^2/(2m)$ is the bare single-particle 
energy and $U_{\bf k}=\sum_{{\bf q}} 
\langle {\bf k},{\bf q}|V|{\bf k},{\bf q}\rangle$ is the 
Hartree-Fock potential. 
The equation for the average number of neutrons is easily 
obtained from Eq. (\ref{def-n}) as 
\beq 
N = 2 \sum_{\bf k} v_{\bf k}^2 \; . 
\label{enne}
\eeq
Finally, from Eq. (\ref{def-n0}) one finds that 
the condensate number of paired neutrons 
is given by \cite{sala-odlro,ortiz}
\beq 
N_{con} = 2 \sum_{\bf k} u_{\bf k}^2 v_{\bf k}^2 \; . 
\label{enne-con}
\eeq 

\begin{figure}[t]
\centerline{\epsfig{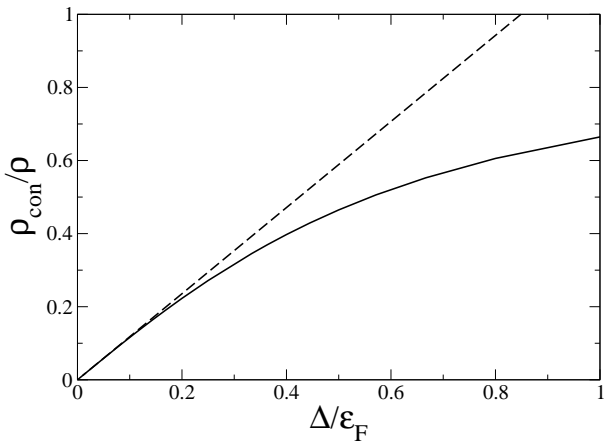}}
\small 
\caption{Condensate fraction $\rho_{con}/\rho$ 
as a function of the scaled pairing gap $\Delta/\epsilon_F$. 
The solid line is obtained by using Eq. (\ref{frac}). 
The dashed line is obtained by using Eq. (\ref{frac-bcs}), 
valid in the regime $\Delta/\epsilon_F\ll 1$.} 
\label{fig1}
\end{figure} 
 
The numerical solution of Eqs. (\ref{delta}), (\ref{enne}), 
and (\ref{enne-con}) is a hard computational task. However, the 
problem can be simplified under some reliable assumptions \cite{fetter}. 
In Eqs. (\ref{uev}) we adopt the approximations 
$\tilde{\epsilon}_{\bf k} \simeq \epsilon_{\bf k}+U_{k_F}$ 
and $\mu \simeq \epsilon_{k_F}+U_{k_F}$, where 
$k_F=(3\pi^2\rho)^{1/3}$ is the Fermi wavenumber of 
the non-interacting Fermi gas with density $\rho$. 
In this way we get 
$\tilde{\epsilon}_{\bf k}-\mu \simeq \hbar^2k^2/(2m) - \epsilon_F$ 
where $\epsilon_F=\epsilon_{k_F}=\hbar^2(3\pi^2\rho)^{2/3}/(2m)$ 
is the Fermi energy of the non-interacting Fermi gas. 
Notice that this is formally equivalent of setting directly 
$\tilde{\epsilon}_{\bf k} \simeq \epsilon_{\bf k}$ 
and $\mu\simeq \epsilon_F$ in Eqs. (\ref{uev}). 
In addition, we set $\Delta_{\bf k} \simeq \Delta_{k_F}=\Delta$. 
Under these assumptions, in the three-dimensional  
continuum limit the gap equation (\ref{delta}) 
of the neutron matter becomes 
\beq 
1 = {1\over 2} \int {d^3{\bf k}\ d^3{\bf r}\over (2\pi)^3} 
{V({\bf r}) \ e^{i({\bf k}-{\bf k}_F)\cdot {\bf r}} 
\over \sqrt{({\hbar^2k^2\over 2m}-\epsilon_F)+\Delta^2}} \; . 
\label{delta-simpler}
\eeq 
Moreover, from the number equation (\ref{enne})  
we find the total density as 
\beq  
\rho = {1\over 2} {(2m)^{3/2} \over 2 \pi^2 \hbar^3} \,
\Delta^{3/2} \, I\left({\epsilon_F \over \Delta}\right)  \, ,  
\label{rho} 
\eeq 
where $I(x)$ is the monotonic function 
\beq 
I(x) = \int_0^{+\infty} y^2 \left( 1 - {y^2-x\over \sqrt{(y^2-x)^2+1}}
\right) dy \; .  
\eeq
In a similar way from Eq. (\ref{enne-con}) we get 
the condensate density of the neutron-neutron pair 
\beq 
\rho_{con} = {m^{3/2} \over 8 \pi \hbar^3} \,
\Delta^{3/2} \sqrt{{\epsilon_F\over \Delta}+
\sqrt{1+{\epsilon_F^2 \over \Delta^2} }} 
\label{rho-con} \; .   
\eeq 
The condensate fraction follows immediately as the ratio between 
$\rho_{con}$, given by Eq. (\ref{rho-con}) and $\rho$, given 
by Eq. (\ref{rho}), namely 
\beq 
{\rho_{con}\over \rho} = {\pi\over 2^{5/2}} {\sqrt{{\epsilon_F\over \Delta}
+\sqrt{1+{\epsilon_F^2\over\Delta^2}}}\over I({\epsilon_F\over \Delta})} 
\label{frac}
\eeq 
Notice that this fraction is expressed only in terms of 
the ratio ${\epsilon_F/\Delta}$. In the deep BCS regime 
where $\Delta/\epsilon_F \ll 1$ then 
$\sqrt{{\epsilon_F/\Delta}+\sqrt{1+{\epsilon_F^2/\Delta^2}}} 
\simeq \sqrt{2} \ \left({\epsilon_F/\Delta}\right)^{1/2}$
and $I\left({\epsilon_F/\Delta}\right) 
\simeq (2/3) \big({\epsilon_F/\Delta}\big)^{3/2}$ 
and consequently the condensate fraction becomes 
\beq 
{\rho_{con}\over \rho} = {3\pi\over 8} {\Delta\over \epsilon_F} \; . 
\label{frac-bcs}
\eeq
In Fig. \ref{fig1} we plot the 
condensate fraction $\rho_{con}/\rho$ 
as a function of the scaled pairing gap $\Delta/\epsilon_F$. 
The solid line is obtained by using Eq. (\ref{frac}), while 
the dashed line is obtained by using Eq. (\ref{frac-bcs}). The figure 
shows that for $\Delta/\epsilon_F \leq 0.2$ the two curves practically 
coincide. 

\begin{figure}[t]
\centerline{\epsfig{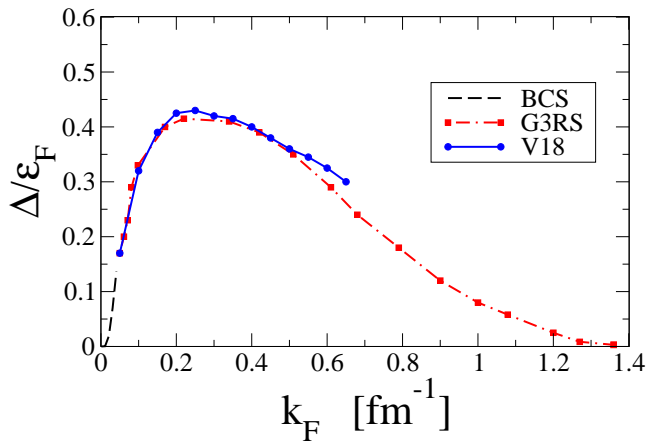}}
\small 
\caption{(Color online). Scaled pairing gap 
$\Delta/\epsilon_F$ of neutron matter 
as a function of the Fermi wave number $k_F$. 
The dashed line is obtained with Eq. (\ref{delta-bcs}); 
the filled squares with the dot-dashed line 
are the results obtained with the $G3RS$ 
nuclear potential \cite{matsuo}; the filled circles with the solid line 
are the results obtained with the Argone 
$V18$ nuclear potential \cite{carlson}.} 
\label{fig2}
\end{figure}

To estimate the condensate fraction $\rho_{con}/\rho$ in neutron 
matter on the basis of Eq. (\ref{frac}) 
it is necessary to obtain $\Delta/\epsilon_F$ from 
Eq. (\ref{delta-simpler}) or, if possible, directly from Eq. (\ref{delta}). 
This problem has been investigated by various authors 
\cite{matsuo,carlson,var1,var2,var3,var4,var5,var6,var7}. 
For very dilute neutron matter, the 
nucleon-nucleon potential $V({\bf r}-{\bf r}')$ in the 
$^{1}S_{0}$ channel can be parametrized in terms of only 
the s-wave scattering length $a_s=-18.5$ fm \cite{fetter}. 
Indeed, in the regime $k_F |a_s|\ll 1$, the pairing gap 
is exponentially small and from Eq. (\ref{delta-simpler}) 
one finds \cite{fetter} 
\beq 
{\Delta\over \epsilon_F} = {8\over e^2} e^{-\pi/(2k_F|a_s|)} \; . 
\label{delta-bcs}
\eeq 
It follows immediately from Eq. (\ref{frac-bcs}) 
that in this regime of very-low density (BCS regime) 
the condensate fraction reads 
\beq 
{\rho_{con}\over \rho} = {3\pi\over e^2} e^{-\pi/(2k_F|a_s|)} \; . 
\label{frac-bcs2}
\eeq 
At higher densities the pairing gap $\Delta$ does not follow an 
exponential grow but instead it reaches a maximum and eventually 
goes to zero \cite{matsuo,carlson,var1,var2,var3,var4,var5,var6,var7}. 
In Fig. \ref{fig2} we report 
the scaled pairing gap $\Delta/\epsilon_F$ 
as a function of the Fermi wave number $k_F$. 
The solid line, obtained with Eq. (\ref{delta-bcs}), shows the 
exponential grow which is reliable only in the very low density 
BCS regime. The filled squares are the results obtained 
by Matsuo \cite{matsuo} by solving Eq. (\ref{delta}) with the 
$G3RS$ nuclear potential \cite{g3rs}; 
the filled circles are the results 
obtained by Gezerliz and Carlson \cite{carlson} 
with the Argone $V18$ nuclear potential \cite{v18}. The figure shows 
that the two nucleon-nucleon potentials give very similar results. 

Instead of solving Eq. (\ref{delta-simpler}) 
we use the data of $\Delta/\epsilon_F$ shown in Fig. \ref{fig2} 
to derive the condensate fraction $\rho_{con}/\rho$ 
by using our Eq. (\ref{frac}). 
It is then physically relevant to plot the obtained condensate fraction
$\rho_{con}/\rho$ as a function of the density $\rho=k_F^3/(3\pi^2)$.

\begin{figure}[t]
\vskip 0.5cm
\centerline{\epsfig{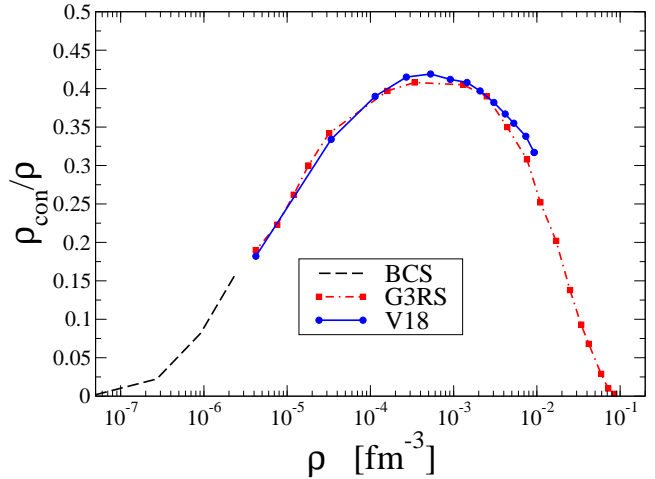}}
\small 
\caption{(Color online). Condensate fraction $\rho_{con}/\rho$ 
of neutron matter as a function of the density $\rho$. 
It is derived by using the data of $\Delta/\epsilon_F$ 
of Fig. \ref{fig2} and Eq. (\ref{frac}).} 
\label{fig3}
\vskip -0.2cm
\end{figure}  

The results are shown in Fig. \ref{fig3}, where the horizontal 
axis is in logarithmic scale. 
At very low density $\rho$ the neutron matter behaves like a quasi-ideal 
Fermi gas with weakly correlated Cooper pairs 
and the condensate fraction $\rho_{con}/\rho$ is 
exponentially small (BCS regime, dashed line). By increasing the density 
$\rho$ the attractive tail of the neutron-neutron potential becomes 
relevant and the condensate fraction $\rho_{con}/\rho$ grows significantly.  
On the basis of the data obtained with the Argone $V18$ potential 
(filled circles with dot-dashed line), the maximum of the condensate 
fraction is $(\rho_{con}/\rho)_{max}= 0.42$ 
at the density $\rho=5.3\cdot 10^{-4}$ fm$^{-3}$ which corresponds to 
the Fermi wave number $k_F= 0.25$ fm$^{-1}$. By further increasing 
the density $\rho$ the repulsive core of the neutron-neutron potential 
plays an important role in destroying the correlation of Cooper pairs 
and the condensate fraction $\rho_{con}/\rho$ slowly goes to zero 
(filled squares with solid line). Remarkably, the results 
of Fig. \ref{fig1} are fully consistent with the Monte Carlo 
value $\rho_{con}/\rho \simeq 0.35$ at $\rho=0.003$ fm$^{-3}$ 
one extracts from the finite-temperature Path Integral 
Monte Carlo data of Wlazlowski and Magierski \cite{magierski1,magierski2}. 

As discussed in the introduction, the properties of low-density 
neutron matter are important for the understanding of neutron 
star crusts \cite{star} and also the exterior of neutron-rich 
nuclei \cite{rich}. Our present investigation gives 
a contribution to this fascinating field of research: 
the condensate fraction of neutron matter reaches its maximum value 
of about $40\%$ in correspondence to the maximum of the pairing gap 
at the density of about $5\cdot 10^{-4}$ fm$^{-3}$, 
which is much smaller than the nuclear saturation 
density $0.16$ fm$^{-3}$. It is important to stress 
that neutron-neutron pairs do not form a true bound state 
and the system cannot reach the unitarity limit (where 
there is the formation of the bound state) nor the deep BEC regime 
characterized by a gas of di-neutrons with 
$100\%$ condensate fraction \cite{matsuo}. Nevertheless, 
the maximum value of the condensate fraction we have found for 
neutron matter is not too far from to the value of about $70\%$ 
calculated for the unitarity Fermi gas \cite{sala-odlro}. Indeed our 
results clearly show a BCS-quasiunitary-BCS crossover 
by increasing the neutron density. 

The author thanks Prof. Masayuki Matsuo for making available his 
numerical data and for useful suggestions.

\end{document}